# Performance Analysis of Traditional and Network Coded Transmission in Multi hop Wireless Networks

[1]Muhammad Ali and Alister Burr

*Abstract*— Infrastructure-less Multi-hop Wireless Networks are the backbone for mission critical communications such as in disaster and battlefield scenarios. However, interference signals in the wireless channel cause losses to transmission in wireless networks resulting in a reduced network throughput and making efficient transmission very challenging. Therefore, techniques to overcome interference and increase transmission efficiency have been a hot area of research for decades. In this paper two methods for transmitting data through infrastructure-less multi hop wireless networks, *Traditional* (TR) and *Network Coded* (NC) transmission are thoroughly examined for scenarios having one or two communication streams in a network. The study has developed network models in MATLAB for each transmission technique and scenario. The simulation results showed that the NC transmission method yielded a better throughput under the same network settings and physical interference. Furthermore, the impact of increasing numbers of hops between source and destination on the network capacity and the communications latency was also observed and conclusions were drawn.

*Index Terms*— Wireless Communication Networks, Infrastructure-less Wireless Networks, Ad-hoc Networks, MANETs, SINR, Transmission Scheduling, Network Coding.

## I. INTRODUCTION

In emergency situations as well as battlefield scenarios, there may be a critical need of high data rate and reliable wireless communications whereas the conventional enabling infrastructure may be unavailable, non-existent or damaged [1].

Due to these challenges, tactical systems require a reliable, robust, and secure communication system. In such situations Infrastructure-less multi hop wireless networks may be able to provide these services. However, not all terminals may be able to establish direct links with one another and hence infrastructure-less multi-hop wireless networks (IMWNs) may be the most suitable solution.

However, since in these networks communication between nodes takes place over radio channels and all nodes use the same frequency band, any node-to-node transmission will add to the level of interference experienced by other users. Variations in network parameters like size (number of nodes), density (relative positions of nodes) and data traffic per node could have strong influence on the interference experienced by nodes throughout the network. It has been established that radio channel capacity decreases as the wanted signal carrier power to interference ratio (C/I) decreases [3]. Therefore, good estimates of interference levels are pivotal for performance evaluation of IMWNs.

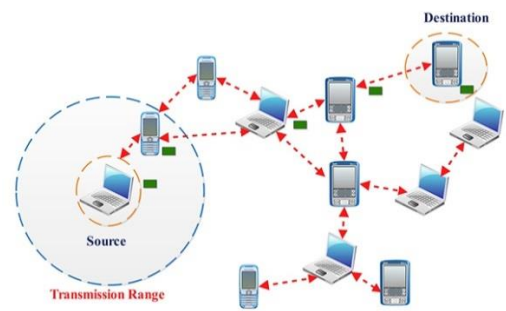

Fig. 1. Wireless Infrastructure less Multi-hop Network [2]

The architecture and functioning of IMWNs have been discussed at length in previously published papers that also highlight the problems that arise in such networks [2-4].

When a receiver in a wireless multi-hop network is out of the coverage/transmission range of the transmitter, relaying is needed. *Routing* facilitates the relaying of data packets through intermediate nodes or *relays*. Hence, efficient routing of data packets in wireless networks is also crucial to the performance of wireless multi-hop network [5–9].

Similarly, when the source and destination nodes are far apart and hence the number of hops needed to connect them is large, multi-hop wireless networks for field monitoring often suffer from frequent packet losses due to attenuation and fading on each link as well as radio interferences of simultaneous transmissions among nodes [18].

Using basic routing algorithms in a wireless environment could lead to problems such as heavy concentration of packets at a node, empty set of neighbours (when using Greedy Forwarding technique), flat addressing, widely distributed information, large power consumption, interference, and load balancing problems [9].

[1] This work was supported by DSTL (Defence Science and Technology Laboratories, U.K.
M. Ali and A. Burr are currently with the School of Physics, Engineering & Technology (PET), University of York, UK. (e-mail: mma601@york.ac.uk)

There are several kinds of wireless networks based on infrastructure-less or Ad-hoc network technology among which the most widely used are *Mobile Ad hoc Networks* (MANETs). MANETs are autonomously self-organized and self-configuring networks without infrastructure support [10]. Others include *Vehicular Ad hoc Networks* (VANETs) and *Wireless Sensor Networks* (WSNs). Mobility and the absence of any fixed infrastructure make MANET very attractive for time-critical applications such as emergency situations and military operations [10].

This paper introduces a simplified model for measuring the impact on the network capacity or throughput of infrastructure-less multi hop networks with both TR and NC transmission with an appropriate transmission scheduling.

Consider, as a simple example, a network containing two terminals that need to exchange data (in both directions) plus a single relay which forwards data from one terminal to the other as illustrated in Fig. 2.

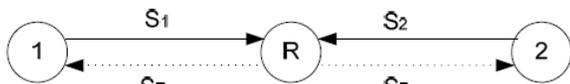

Fig. 2. Network Coded Transmission [11]

In TR transmission the packets arriving at the relays are simply cached and then forwarded in the next timeslot to the next node in the network. In this case a total of four timeslots are needed to exchange two packets, one in each direction [11].

On the other hand, in NC transmission each node transmits to the relays in two successive timeslots data packets $S_1$ and $S_2$ respectively. At the relay an algebraic operation is performed on the two data packets $S_1$ and $S_2$ (usually they are combined using a bitwise XOR operation) [23].

In the third timeslot, the resulting coded data packet $S_R$ is transmitted back to both parent nodes. Given each node itself transmitted a data packet can extract the data packet from the other node by reversing the algebraic operation and thereby completing the data exchange between the two nodes. Hence, NC requires relatively fewer timeslots than TR (three rather than four) to transfer the two data packets increasing throughput per timeslot since more transmissions can occur in same time interval [11]. The same principle may be extended to connections involving more than two hops, as discussed in section IIIA.

To determine the network capacity, the network model first calculates the signal to interference plus noise ratio (*SINR*) at a node by taking in account total number of nodes available to transmit, hop length, number of hops, traffic at relay nodes, transmission power, and interference levels at nodes. For both types of transmissions, bi-directional traffic in half-duplex mode [13] was assumed.

The network model also requires the transmission scheduling period, path loss exponent, type of transmission used, and transmitter coverage distance (and hence maximum allowed distance between nodes), to determine the *SINR* and thereafter compute the throughput per node using the Shannon formula.

The investigation carried out showed how changes made in certain network parameters caused variation in the interference experienced at nodes and hence the network capacity.

The model was also used to validate whether further improvement in throughput could be achieved through appropriate selection of a transmission scheduling period which might reduce the interference from simultaneous transmitters.

This paper is organised as follows. In section II, the network model is described in terms of its physical characteristics and the key assumptions made. In section III, the functioning of the network model is explained. In section IV, the process for computing the received *SINR* levels have been explained which involves calculating the signal power received at a node while identifying and quantifying the causes of signal degradation. In section V, the network capacity or throughput is calculated from simulating the network model for several different settings for the parameter values. In section VI, results obtained from simulations are critically evaluated. Impact on the network's throughput of varying a certain parameter while keeping other parameters constant is thoroughly examined. The conclusions drawn are summarised in section VII.

II. MODEL DESCRIPTION

In this section the network model used to evaluate impact of employing NC transmission on the network capacity in an IMWN will be described. The conditions assumed for the radio propagation environment, and the assumptions made with regards to transmission will also be explained. As suggested earlier, it is assumed that pairs of terminals wish to exchange information with one another on an ad-hoc basis, via intermediate relay nodes: in the sequel we will refer to such a pair of terminals and the associated relay nodes as a *communication stream*. We note that there may be several such communication streams in the network which are simultaneously active.

*A. Network Layout*

The layout for this IMWN is designed to support one or two coexisting communications streams/routes comprising nodes that are aligned and at mutually equal spacings. The number of nodes per stream is $N_o$ and the total number of nodes in the network is $N = 2N_o$, for the case of two streams. There are always equal numbers of nodes per stream arranged in two rows for two communication streams. Hence by setting the total number of nodes *N* and maximum length of a hop between a pair of adjacent nodes, the maximum length of a route can be determined and vice versa. This layout allows the network model to determine the precise location of individual nodes in the network.

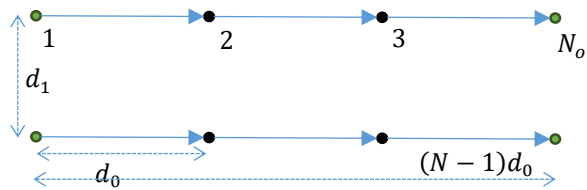

Fig. 3. Network Layout with two communication streams

In Fig. 3, the nodes in each communication stream are spaced at equal distances $d_0$ from each other in a row hence giving a maximum route length of $(N_o - 1)d_0$. Further, the rows of nodes are at a distance $d_1$ apart from each other and varying $d_1$ controls the interference level. Also, the number of nodes in each stream or route can be varied by changing the source and/or destination node.

The simplicity of this layout allows the network model to find a route for a communication stream with less complexity. In the one communication stream case, the total number of nodes given to the network model must be at least 3 to account for one communication streams formed of one pair of source and destination nodes and a relay node, to facilitate multi-hopping. Whereas in the two communication streams case, the number of nodes given to the network model must be at least 6, to account for two communication streams. In this paper the total number of network nodes is restricted to be less than 7 for the one communication stream case and 14 for the two communication streams case.

The reason for defining the highly simplified layout containing just two parallel streams with uniform hop lengths illustrated in Fig. 3 is as follows. The throughput of a IMWN is primarily dependent on the signal strength and interference experienced at each node. Interference occurs between different hops of the same communication stream as well as between different streams. The exact signal and interference levels are clearly dependent on the specific layout of the nodes, however, in this paper the effects of NC versus TR transmission and the effect of transmission scheduling on the network capacity of an IMWN is the prime focus of interest, and hence we define a very simple model which includes interference and accounts for the effect of scheduling on it.

### B. Application of Scheduling to transmitters

*Scheduling* [22] is a method by which in a certain timeslot only specific nodes may transmit. Conversely each node will be assigned a timeslot in which it may transmit. Here we assume a periodic *schedule* with period $Z$ timeslots:

That is, every node transmits according to a *schedule* which repeats every $Z$ timeslots. In the layout defined in section IIA we assume that nodes are numbered sequentially within a stream, as shown in Fig. 3. Then if we consider the forward direction only, we may also assume that in timeslot $i$ node $i$ transmits to node $i + 1$.

This also implies that in timeslot $i$, node $i + Z$ also transmits to node $i + Z + 1$, if such a node exists i.e. if $i + Z + 1 \leq N$, and hence $i + Z \leq N - 1$. (Also, node $i + 2Z, i + 3Z$ etc may transmit if they and the nodes they would transmit to exist). Hence *intra-stream interference* may occur to node $i + 1$.

A minimum scheduling period of $Z = 2$ is needed in any case because of the *half-duplex constraint*: that is, a node cannot normally transmit and receive simultaneously [13], because its transmission would cause very serious interference to its own receiver. Hence a maximum of half the nodes may transmit at a time.

This communication stream may also be subject to interference from a second coexisting communication stream with identical number of nodes and mutual spacing $d_0$ and at a distance $d_1$ apart from the communication stream of interest. Both streams operate according to a schedule with the same period, and it is assumed that the slot periods across the network are synchronised (and neglect any propagation delay).

### C. Link Modelling Assumptions

The main assumptions forming the basis for this model and the calculation of *SINR* leading to the network data rate (Capacity) are given in the following.

1) All network nodes are stationary

This means that the distances within the network remain constant which simplifies the calculation of signal and interference.

2) Path loss follows inverse power law

Calculations for the signal and interference power received at nodes in this paper are based on a generic inverse power law [12] or radio wave transmission. The mean received signal power $P_{RX}$ (in Watts) can be represented as:

$$P_{RX} = P_{TX} K \left(\frac{d_o}{d}\right)^\eta \qquad (1)$$

where $K$ is a constant such that:
$$K = G_{TX} G_{RX} \left(\frac{\lambda}{4\pi d_o}\right)^2$$
and:
$P_{Tx}$ is input power at the transmitting antenna input terminals
$P_{Rx}$ is power available at the receiving antenna Output terminals
$d_0$ is the reference distance taken as 1 m
$d$ is the distance between the receiving and transmitting antennas
$\lambda$ is the wavelength of the *radio frequency* used
$\eta$ is the path loss exponent

The *Path Loss Exponent* (PLE) denoted by $\eta$ is a critical parameter when calculating the power received at the receiver node. $\eta$ values are based on measurements obtained in different mobile radio environments generally range from 2 for free space to 6 in buildings [16]. Therefore, the variation in η depends on the radio propagation environment in terms of how much cluttered with objects that are causing the signal to reflect back or interfere. Higher values for η indicate faster decay of radio signals. (1) is defined in such a way that different values of $\eta$ can be chosen, but for $\eta = 2$ the formula agrees with the Friis law [13]. For this model the environment was chosen to be comprising many reflecting surfaces and obstacles and hence a PLE value of 4 was used in received power calculations.

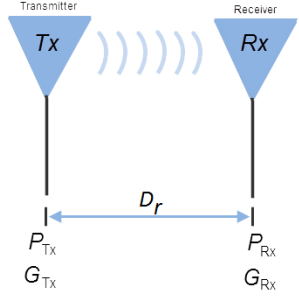

Fig. 4. Communications link between a node pair [14]

The coverage area of a node was defined as a circular area around this node with a *radius* equal to the *Coverage Range* specified in this model. A node can have direct communication will all nodes that fall within its coverage area. This model is relatively and does not account for small- and large-scale fading variations around signal's mean power [15].

3) The receiver may be modelled based on noise figure

The *Noise Power* $P_n$ might be calculated as:

$$P_n = FkTB \qquad (3)$$

where:

$k$ is Boltzman's Constant (1.38×10$^{-23}$ J K$^{-1}$)
$T$ is room temperature (300K)
$F$ is noise factor or noise figure
$B$ is signal bandwidth

4) Link capacity may be estimated using the Shannon bound

The most significant assumption made by the network model with regards to calculating the throughput or network capacity at the nodes in a communication stream is that it is bounded by the Shannon–Hartley theorem. The network model estimates the network capacity based on this bound [3] which states that for a communication channel the channel capacity $C$, meaning the theoretical maximum upper bound on the information rate of data that can be communicated at an arbitrarily low error rate using an average received signal power $S'$ through an analogue communication channel depending on the *Additive White Gaussian Noise* (AWGN) of power $P_n$:

$$C = B \log_2[\,1 + SINR\,] \qquad (4)$$

where:

$C$ is the channel capacity in bits per second, a theoretical upper bound on the net bit rate;
$B$ is the bandwidth (in Hertz) of the channel;
$SINR$ s the signal-to-interference plus noise ratio [3].

## III. SIMULATION METHODOLOGY

### A. Simulator operation

The key parameters are listed in table 1, along with the range of values investigated and the default values assumed.

| Parameter Name | Denoted in Model | Possible Set of Values | Used Values |
|---|---|---|---|
| Total Number of Nodes (in Network) | $N$ | $6 < N < 14$ | 10 |
| Hop Length (distance between adjacent nodes) | $d_0$ | (100 - 300) m | 100 m |
| Distance between adjacent routes | $d_1$ | (300 - 1000) m | 300m |
| Scheduling period | $Z$ | $Z < N$ | 2-5 |
| Operational Frequency | $f$ | 1- 3 GHz | 2 GHz |
| Path Loss Exponent | $\eta$ | 2 - 4 | 4 |
| Transmission Power | $P_t$ | 1mW – 100mW | 100mW |
| Antenna Gain | $G$ | 1–2 | 1 (0 dB) |
| Noise Factor | $F$ | 2dB – 6dB | 4dB |

Table. 1. Network parameters and their ranges

The user first assigns parameter values and initiates the simulation. The network model then calculates the positions of the given number of nodes in two straight rows of nodes, a distance $d_0$ apart, parallel to each other and separated by a distance $d_1$ and creates a database of their locations.

Using knowledge of node locations, the network model calculates distances between each possible pair of nodes in the network and places them in a matrix. The network model then returns a schematic showing locations of all the nodes in the network. It is assumed that the source and destinations are at the ends of the rows of nodes in this layout.

Since the nodes are aligned in two rows with a fixed mutual distance between nodes in each row, the route between the source and destination and the relay nodes can easily be determined by the network model.

The network model first determines the sets of nodes that provide routes for one or both communication streams as appropriate. The network model then computes the *SINR* at each node of the routes, depending on the transmission scheduling applied, by working out the signal power received at the node and the interference plus environmental noise occurring there.

Assuming that one node in each pair is a transmitter and the other is a receiver and the transmitted power is as given in the parameters table, the received power can be estimated using (1).

After calculating the received power by each node from each node in the network, the intended power and interference received can be worked out. Interference is the sum of unintended signal powers received at a node regardless of the allowed range for reception.

Note that interference is experienced by nodes in each communication stream due to simultaneous transmissions from other nodes in the same stream and from the interfering stream. Note also that the larger the schedule period the fewer nodes will transmit simultaneously, reducing the interference and hence increasing the *SINR* at each receiving node and leading to a higher link throughput at these nodes.

The schedule cycle is applied to all transmissions in the network, that is to both directions of transmission for all communication streams existing in the network.

Once the interference and signal power at a node is calculated, and also the noise power using (3), the *SINR* at that node can be calculated using:

$$SINR = \frac{P_{Rx}}{I_x + P_n} \quad (6)$$

After calculating the *SINR* at the nodes forming the route of the communication stream, the network model applies the Shannon-Hartley formula to calculate the maximum throughput at each node in bits per second (bps).

The lowest data rate found at any node on the route in each direction is the overall bit rate in the that direction. The sum of the two data rates in each direction, divided by the schedule period $Z$ is the capacity per timeslot of this communication stream for the NC case whereas for the TR case it is divided by $2Z$. For a network comprising a single communication stream, this is the overall network capacity.

The total number of nodes in a network limits the number of communication streams that can co-exist. For this project the network model was designed and tested for only up to two simultaneous communication streams with less than 7 nodes each.

In addition to the overall throughput per stream, the modelling software can also provide graphs of the network showing the locations of all nodes, the links in the communication streams and bar charts of the *SINR*s and bit rates per node, but these are not included in this paper Another option is to provide the network model with node IDs for the assumed sources and destinations and give it a range of values for a certain parameter to see the impact on the network capacity due to variation in that parameter. This is the scenario for the results provided and discussed in this paper.

Once the simulations are complete, the network model returns a bar chart showing the overall network capacity results for groups of simulations.

B.  Schedule for TR transmission

A schedule defines the set of nodes that is permitted to transmit in each timeslot of the schedule: we define the set permitted to transmit in timeslot $i, 1 \leq i \leq Z$ as $T(i)$. Assuming a simple forward sequential schedule we may write:

$$T_F(i) = \left\{i + nZ, n \in 0,1 \ldots \left\lfloor \frac{N_o - 1 - i}{Z} \right\rfloor\right\} \quad (7)$$

where $\lfloor x \rfloor$ denotes the floor function, which returns the next integer less than the argument $x$. Then nodes $i, i + Z, i + 2Z \ldots$ transmit in slot $i$ provided these nodes exist and are not the last node in the stream. We assume that when node $i$ transmits a packet, node $i + 1$ receives and stores that packet until its next opportunity to transmit in the forward direction – which is usually the next timeslot, but this may not always be the case.

Similarly, the simple reverse sequential schedule is defined by:

$$T_R(i) = \left\{N_o - i + 1 - nZ, n \in 0,1 \ldots \left\lfloor \frac{N_o - 1 - i}{Z} \right\rfloor\right\} \quad (8)$$

That is, nodes $N_o + 1 - i, N_o + 1 - i - Z, N_o + 1 - i - 2Z \ldots$ transmit in slot $i$ provided these nodes exist and are not the first node in the stream. We assume that when node $i$ transmits a packet, node $i - 1$ receives and stores that packet until its next opportunity to transmit in the reverse direction: again, this is usually but not always the next timeslot.

We assume that the communication stream alternates between the forward and reverse schedules. This means that at the end of a forward schedule period the nodes must store the packets being sent in the forward direction until the next forward schedule period, and similarly for the reverse direction. Figure 6 illustrates the operation of the schedule for $N_o = 5$ and $Z = 3$. The figure shows the progress of packets along the stream in successive timeslots.

An important performance measure for the system is its end-to-end *latency L*: the delay (measured in timeslots) between transmission of a packet from the source node and its arrival at the destination. If $Z \geq N_o - 1$, then the packet is delivered in one schedule period, with one timeslot delay per hop, and hence $L = N_o - 1$. However, if $Z < N_o - 1$, a packet transmitted in the forward direction will need to be buffered at node $Z + 1$ during the reverse part of the schedule, adding $Z$ additional timeslots delay, and similarly for a packet in the reverse direction. In general, if $N_o > nZ + 1$ it will be delayed by $nZ$ timeslots, and hence total latency:

$$L = N_o - 1 + Z \left\lfloor \frac{N_o - 2}{Z} \right\rfloor \quad (9)$$

C.  Scheduling and coding for NC transmission

In NC transmission packets can be transmitted in both directions simultaneously, since network coding can exploit the signals received at both neighbours of a transmitting node. However, scheduling is still necessary to control interference. In this paper we assume that the simple sequential forward schedule is used, as described in section IIIB above.

Network coding in a multi-hop scenario is more complicated than the simple 2 hop example illustrated in Fig 3 above, since there are multiple relays, and multiple packets in transit at any time. The fundamental principle, however, is that a relay node may perform the XOR operation on packets received from the left (that is, travelling in the forward direction along the communication stream) and from the right (i.e. those travelling in the reverse direction). Hence the packets transmitted may be formed from the bitwise XOR combination of several source packets: we assume that each of these packets has a header which indicates which source packets have been combined. Then at each timeslot a relay may decide whether to XOR the packet received in that timeslot with a packet stored from a previous timeslot, or to overwrite the stored packet with the received one.

Note that for any packets $A, B, C$:

$$A \oplus A = B \oplus B = C \oplus C = 0; 0 \oplus A = A \quad (10)$$

where $\oplus$ denotes the bit-wise XOR operation performed on two packets. (It is also equivalent to the bitwise modulo-2 sum). Note that $\oplus$ is commutative and associative, i.e:

$$A \oplus B = B \oplus A; (A \oplus B) \oplus C = A \oplus (B \oplus C) \quad (11)$$

Hence:
$$A \oplus B \oplus B = A \oplus B \oplus C \oplus B \oplus C = A \quad (12)$$

This means that appropriate choice of whether to apply the XOR operation can ensure that the desired packets are transmitted in the desired direction.

Figure 7 illustrates the operation of scheduling and network coding for $N_o = 5, Z = 4$. Note that nodes are numbered from left to right, and arrows indicate transmission of a packet to neighbouring relay nodes. Observe that at the end of timeslot (TS) 3 node 4 XORs packet A received from node 3 with packet B already stored in the node: this ensures that packet B is transmitted to the left, while node 5 (the destination for packet A) receives $A \oplus B$, from which packet A can be recovered using the stored packet B, since $A \oplus B \oplus B = A$. In TS 4 node 3 receives $A \oplus B$ from node 4, and combines it with packet A already stored there, thus regenerating packet B which can be stored there until it can be transmitted to node 2 in TS 7, and finally to node 1 in TS 10.

The schedule means that new packets are transmitted from each source node every $Z$ timeslots, and in the steady state the figure shows that packets are also successfully received at their destinations every $Z$ timeslots. The figure also shows that packets travelling from left to right take $4 = N_o - 1$ timeslots to reach their destination, since they are travelling in the same direction as the schedule: in general, the latency in the forward direction is:

$$L = N_o - 1 \quad (13)$$

However, in the reverse direction they take 10 timeslots: this is because they progress by only one hop every schedule period. In the first TS of its journey (TS 1) it is transmitted to node $N_s - 1$; it waits $Z - 1$ TSs at each of nodes $N_o - 2, N_o - 3$, etc, and hence arrives at its destination after $(N_o - 2)(Z - 1) + 1$ TSs, i.e. in the reverse direction latency is:

$$L = (N_o - 2)(Z - 1) + 1 \quad (14)$$

### D. Scenarios to illustrate scheduling

In the scenario that route of a communication stream comprises more than 3 nodes, there would be multiple relay nodes.

In the case of TR transmission, data packets travel in only one direction in a timeslot i.e. from their source node towards their destination node. Whereas in the case of NC transmission, data packets travel in both directions in a timeslot. Hence fewer timeslots are required to enable an exchange of data packets. Both cases are illustrated below.

At any given instant, NC occurs at the relay nodes that have already received packets in previous instants from two different nodes. With more relay nodes present, larger number of NC operations may take place at a given instant in time.

Now to illustrate this, consider an IMWN where *steady state* has been reached. That is, there are data packets already present at all transmitting nodes of the communication stream in a specific timeslot. For this network, the effect of scheduling when employing either TR or NC based transmission is to be determined.

For Scenario 1, traditional transmission is employed with a scheduling period $Z = 3$ and the total number of nodes in a route for the communication stream is $N = 5$. Assume, in the first timeslot, node 1 (source) from the left to right transmits packet A to node 2 while node 4 (relay 3) transmits to node 5 (destination) some previous packet. In timeslot 2, only node 2 (relay 1) transmits to node 3 (relay 2) and in timeslot 3, only node 3 transmits to node 4 (relay 3). In timeslot 4, node 5 (source) transmits packet B from right to left (which is the reverse movement of packets) to node 4 and node 2 transmits to node 1 (destination) some previous packet. In timeslot 5, only node 4 transmits to node 3 and in timeslot 6, only node 3 transmits to node 2. In timeslot 7, node 1 (source) from the left to right transmits packet C to node 2 while node 4 (relay 3) transmits packet A to node 5 (destination) and this way packet A which started from node 1 has reached node 5. In timeslot 8, only node 2 transmits to node 3 and in timeslot 9, only node 3 transmits to node 4. In timeslot 10, node 5 (source) transmits packet D from right to left to node 4 and node 2 transmits packet B to node 1 (destination) and in this way packet B has reached node 1.

This can be seen in Fig. 6 below; a constant transmission pattern is repeated after every 6 (2*Z) timeslots.

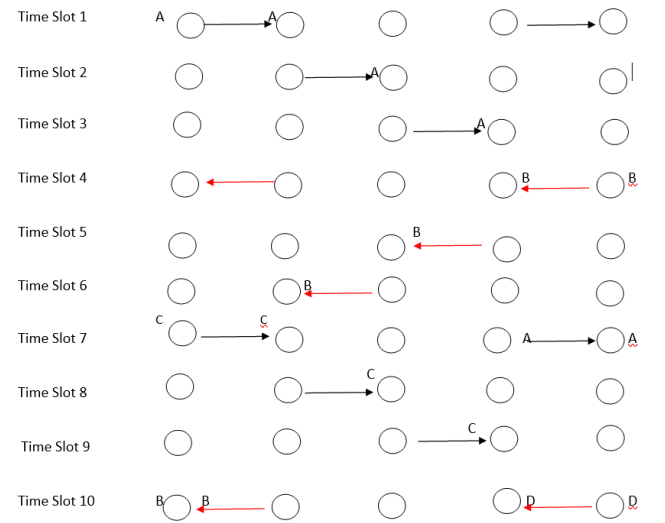

Fig. 6. Scenario 1: scheduling in IMWN with TR

In this scenario, it takes 10 timeslots to complete the exchange of data packets (i.e. 2 data packets) between the two source/ destination nodes (i.e. between node 1 and node 5 respectively). However, on average 2 packets arrive at their respective destinations every $2Z = 6$ timeslots, so the throughput is the maximum rate per communication stream multiplied by a factor $2/6 = 1/3 = 1/Z$.

For Scenario 2, where the IMWN is employing NC transmission with a scheduling period $Z = 4$ and the total number of nodes in a route for the communication stream is $N = 5$. Assume that the data exchange is considered at an initial state where there are no data packets already present at any node in the communication stream.

In timeslot 1, node 1 transmits packet A to node 2 (relay 1) and while node 5 transmits packet B to node 4 (relay 3). In timeslot 2, only node 2 transmits to node 3 (relay 2) and in timeslot 3, node 3 transmits to node 4. Then, since a data packet was already received at node 4 in the previous timeslot, NC occurs at node 4 combining packets A and B as $A \oplus B$. In timeslot 4, node 4 transmits the NC packet $A \oplus B$ to both node and node 5. At node 5, $A \oplus B$ is combined with packet B to form $A \oplus B \oplus B = A$, and hence packet A arrives at node 5. At node 3, where packet A is available, it is combined with $A \oplus B$ to form $A \oplus A \oplus B = B$. In timeslot 5, nodes 1 and 5 transmit again, sending two new packets, C and D to nodes 2 and 4 respectively. In timeslot 6, after node 2 transmits packet C to node 3, since packet B is present from the previous timeslot, NC occurs forming $C \oplus B$. The process then proceeds in a similar way as illustrated in Fig. 7. Hence a data packet 'A' is transmitted (L-R) from node 1 (its source) to node 5 (its destination) in 4 timeslots. On the other hand, a data packet 'B' is transmitted (R-L) from node 5 (its source) to node 1 (its destination) in 10 timeslots. It is seen that in timeslot 4, as the initial data packet from node 1 has reached node 5, steady state is achieved and a constant pattern for transmission is repeated after timeslot 4.

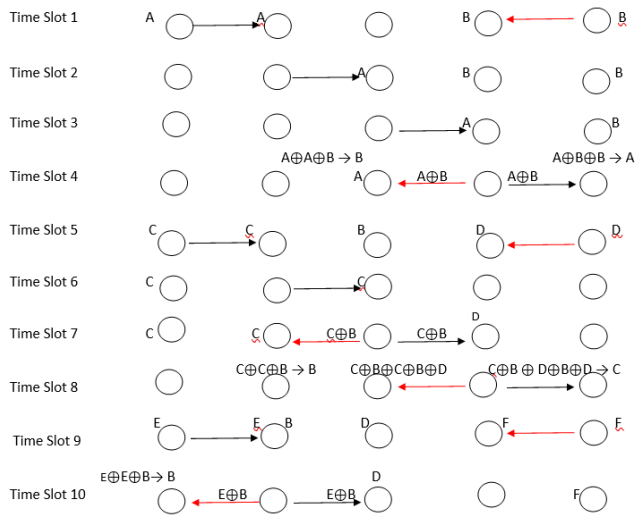

Fig. 7. Scenario 2: scheduling in IMWN with NC

In this scenario, it takes 10 timeslots to deliver the first packet from node 5 to node 1 (right to left). However, in steady state one packet is delivered in each direction per schedule period, so two packets reach their respective destinations in 4 timeslots. Hence the rate is 2/Z which will be multiplied with the data rate calculated for each communication stream to find the actual throughput.

## V. EFFECT ON NETWORK CAPACITY OF VARYING SCHEDULING PERIOD

The impact on the throughput of an IMWN of variation in scheduling period $Z$ was measured using the MATLAB network model developed for this study.

Two scenarios were considered: a single communication stream and two coexisting communication streams in an IMWN with TR and NC transmission employed while taking in account the interference from all simultaneously transmitting nodes.

| Parameter Name | Parameter Denotation | Parameter Values |
|---|---|---|
| Total No. of Network Nodes (One Route) | $N = N_o$ | 6 |
| Total No. of Network Nodes (Two Routes) | $N = 2N_o$ | 12 |
| Hop Length (distance between adjacent nodes) | $d$ | 100m |
| Distance between adjacent routes (Two Routes) | $d_1$ | 300m |
| Transmission power | $P_t$ | 100 mW |
| Bandwidth | $B$ | 1 MHz |
| Scheduling period | $Z$ | 2 - 5 |
| Path loss exponent | $\eta$ | 4 |

Table. 2. Parameter values for one and two route networks

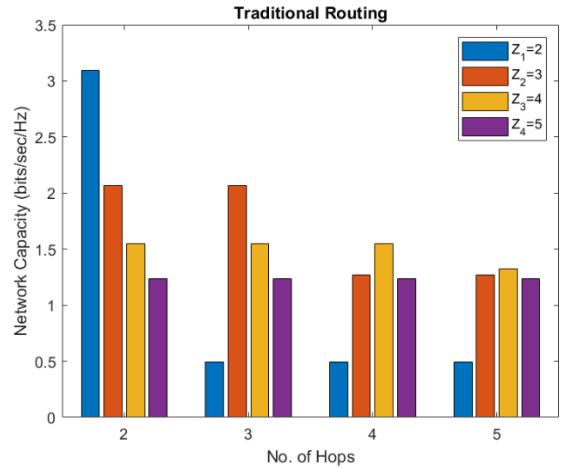

Fig. 6. The network capacity calculated for a single route network with TR for different scheduling periods.

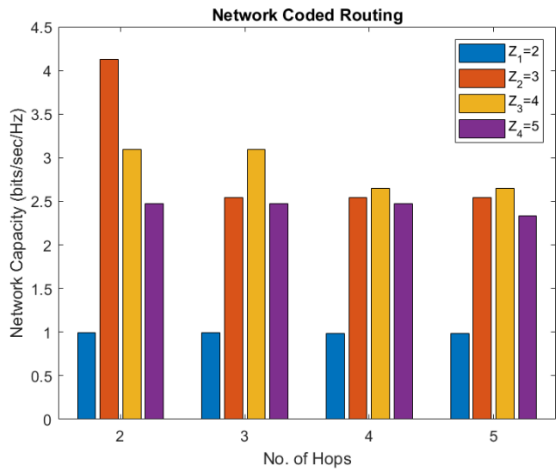

Fig. 7. The network capacity calculated for a single route network with NC for different scheduling periods.

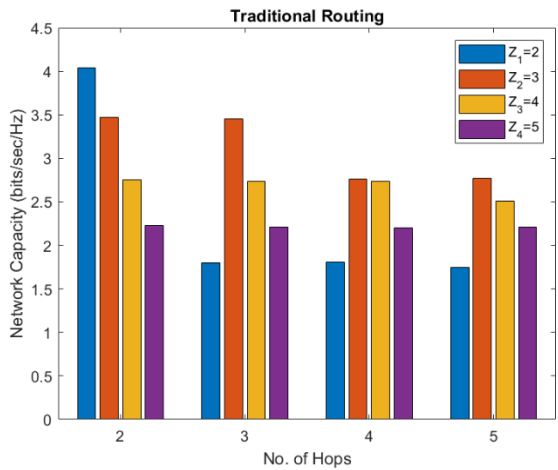

Fig. 8. The network capacity calculated for a two routes network with TR for different scheduling periods.

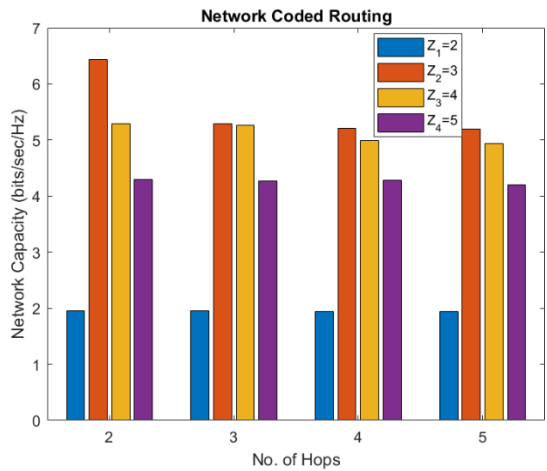

Fig. 9. The network capacity calculated for a two routes network with NC for different scheduling periods.

A. Observations from the simulation results

The results shown for an IMWN comprising a single communication stream indicate that the optimum bit rate for the NC case was obtained when $Z = 4$ and for the TR transmission case when $Z = 3$ for up to 3 hops (optimum is $Z = 2$ for 2 hops) and $Z = 4$ for hops > 3 as seen in Fig. 6 and 7. Data rates for NC transmission were higher than TR transmission data rates in all instances by between 50% and 100% as seen in Table 4.

On the other hand, the results shown for an IMWN comprising two coexisting communication streams, indicate that the optimum data rates for the NC and the TR transmission case were obtained when $Z = 3$ for any number of hops (except for 2 hops in the TR case) as seen in Fig. 8 and 9. Again, data rates for NC transmission were higher than TR transmission data rates in all instances, by between 52% and 88% as seen in Table 4.

This result is due to application of a higher scheduling period i.e. fewer nodes allowed to transmit simultaneously, hence fewer network transmissions in both communication routes and lesser interference generated. This resulted in a higher *SINR* at a receiving node and therefore higher data rates at nodes resulting in an improved network capacity.

The data rates calculated by the network model have shown that the data rate depends on number of hops in a route as well as 'Z' the scheduling period. In all instances, IMWNs with NC transmission display a higher Shannon capacity bound (between 50% to 100%) than those with TR transmission. Furthermore, the data rates at receiving nodes are affected by interferences due to simultaneous transmissions in their vicinity. These data rates indicate that IMWNs with NC transmission are less affected when there are multiple communication streams co-existing as compared to those with TR transmission.

| No. of Nodes $N$ | No. of Hops $N-1$ | Transmission Scheme | Optimum Scheduling period 'Z' | | Data Rate 'C' Shannon's Eq. | | $\frac{C_{NC} - C_{TR}}{C_{TR}}\%$ OS | $\frac{C_{NC} - C_{TR}}{C_{TR}}\%$ TS | $\frac{C_{OS} - C_{TS}}{C_{TS}}\%$ |
|---|---|---|---|---|---|---|---|---|---|
| | | | OS | TS | OS | TS | | | |
| 3 | 2 | TR | 3 | 3 | 2.064 | 1.749 | 50% | 85% | -15% |
| | | NC | 4 | 3 | 3.095 | 3.234 | | | +4% |
| 4 | 3 | TR | 3 | 3 | 2.064 | 1.742 | 50% | 52% | -15% |
| | | NC | 4 | 3 | 3.095 | 2.654 | | | -14% |
| 5 | 4 | TR | 4 | 3 | 1.547 | 1.390 | 71% | 88% | -10% |
| | | NC | 4 | 3 | 2.645 | 2.608 | | | -2% |
| 6 | 5 | TR | 4 | 3 | 1.322 | 1.390 | 100% | 87% | +5% |
| | | NC | 4 | 3 | 2.645 | 2.598 | | | -2% |

Table. 4. Comparing data rates between TR and NC transmission networks and for one stream (OS) and two streams (TS) cases

## VI. DISCUSSION

The prime objective of this study was to evaluate the impact on the network capacity of ad-hoc networks when employing TR and NC for transmitting data packets across the network in the presence of interference from simultaneously transmitting nodes. This study verifies that interference levels have a profound effect on the SINR achieved at relay nodes and hence on the network capacity. However, there are other parameters such as network density or hop length, path loss exponent, scheduling period applied, the number of co-existing communication streams and their mutual distance, that also affected the network capacity. The simulation results presented in this paper are based on the impact of varying the scheduling period applied to transmitting nodes while keeping the other parameters constant for a network with one or two communication streams when TR or NC transmission is employed.

The overall results obtained from the network models indicate that the interference levels in IMWNs result not only from new generated traffic per node, but also from relay traffic that is hopping from source to destination throughout the network. The number of hops from any source to any other destination, determine the amount of relay traffic in such networks.

While comparing TR to NC transmission in terms of the throughput performance, in all the simulations carried out the network capacity was found to be substantially higher when the IMWN had NC transmission employed as compared to TR transmission under the same conditions applied. These results were independent of whether the IMWN comprised of one or two communication streams as seen in Table. 4.

Further, throughput result obtained for an IMWN with one communication stream was better than that with two communication streams, due to the additional interference. This was true for an IMWN employing the TR or the NC transmission technique as evident from the results in Table. 4. Having more than one communication stream present in the infrastructure-less multi-hop network had an adverse effect on the overall network capacity. This indicated that the network capacity was expected to further fall as the number of communication streams coexisting in the network increased.

## VII. CONCLUSION

The model discussed in this paper determines the network capacity of an IMWN by calculating the SINRs and throughput at nodes at given conditions while using various transmission techniques to improve network efficiency. The network model provides an insight to how network capacity varies with respect to increase in the communication streams within the IMWN.

The study presented in this paper also offers experimental evidence of improvement in the overall network capacity through application of interference counter measures specifically transmission scheduling to the transmitting nodes.

To conclude, this network model provided a practical estimation of the interference levels experienced at nodes, effectiveness of applying transmission scheduling, and a baseline to measure the impact of employing various transmission techniques to reduces signal losses and improve the throughput in IMWNs. However, the data rates drop significantly as the number of (inter-node) hops between the source and destination increase.

### A. Future Work

The next step would be to continue this research work by developing network models for IMWN with network nodes that are *randomly* placed and may or may not be stationary. In this case, finding the optimum route between the source and destination nodes would no longer be as straightforward as it was in the IMWN with an aligned nodes layout described in this paper. Therefore, it would be necessary to employ a *Shortest Path Spanning Tree Algorithm* (SPST) to determine the *least cost* route between source and destination nodes by assigning *weights* to links between each pair of nodes in the network.

Furthermore, network models that represent the *Physical Layer Network Coded* (PNC) transmission technique could also be developed. This would facilitate evaluation of the throughput performance and comparison of TR, NC and PNC transmission techniques for use in IMWNs.